\documentclass[prl,tightenlines]{revtex4}
\usepackage{amsmath}
\usepackage{amssymb}
\usepackage{amsfonts}
\newcommand{\beq}{\begin{equation}}
\newcommand{\eeq}{\end{equation}}

\newcommand{\p}{\partial}
\newcommand{\mc}[1]{\mathcal{#1}}
\newcommand{\md}{\mathcal{D}}
\newcommand{\wt}{\widetilde}
\newcommand{\ov}{\overline}

\newcommand{\GeV}{{\rm GeV}}

\newcommand{\ha}{\frac{1}{2}}
\def\slashed#1{\setbox0=\hbox{$#1$}             
   \dimen0=\wd0                                 
   \setbox1=\hbox{/} \dimen1=\wd1               
   \ifdim\dimen0>\dimen1                        
      \rlap{\hbox to \dimen0{\hfil/\hfil}}      
      #1                                        
   \else                                        
      \rlap{\hbox to \dimen1{\hfil$#1$\hfil}}   
      /                                         
   \fi}                                        %

\newcommand{\LUV}{\Lambda_{\rm UV}}

\begin{document}

\renewcommand{\thefootnote}{\fnsymbol{footnote}}

\setcounter{page}{1}



\title{Low-energy constraints on $\kappa$-Minkowski extension of
the Standard Model}

\author{Pavel A. Bolokhov$^{\,(a,b)}$, Maxim Pospelov$^{\,(c,d)}$}

\affiliation{
$^{\,(a)}${\it Physics and Astronomy Department, University of Pittsburgh, Pittsburgh, Pennsylvania, 15260, USA}\\
$^{\,(b)}${\it Theoretical Physics Department, St.Petersburg State University, Ulyanovskaya~1,
        Peterhof, St.Petersburg, 198504, Russia}\\
$^{\,(c)}${\it Department of Physics and Astronomy, University of Victoria, 
     Victoria, BC, V8P 1A1 Canada}\\
$^{\,(d)}${\it Perimeter Institute for Theoretical Physics, Waterloo,
Ontario N2J 2W9, Canada}
}

\begin{abstract}
We investigate the phenomenological consequences of $\kappa$--Minkowski extension of the 
Standard Model, working in the linear order in inverse $\kappa$. At this order the
$*$-deformed Lagrangian can be expanded in the series of dimension five operators that have non-trivial 
transformation properties under the ordinary Lorentz invariance. 
Such operators cause the Lorentz-violating signatures at low energies, and in particular 
lead to the anomalous spin precession linked to the external direction. The experimental 
bounds on this phenomenon then restrict parameter $\kappa$ to be above $10^{23}$ GeV,
making it difficult to impose a direct connection between this theory and quantum gravity. 
\end{abstract}

\maketitle

\section{Introduction}

Noncommutative Field Theory has become a focus of extensive theoretical studies during the 
recent years.
Following the work of Connes, Douglas and Schwarz \cite{Connes:1997cr} and the later paper
by Seiberg and Witten \cite{Seiberg:1999vs}, a tremendous amount of attention
was brought to various aspects of $ \theta $-noncommutativity arising 
as a certain limit of the string theory. 
The noncommutative space-time is given by the algebra of the noncommutative coordinates, which in the most
general form is \cite{Lukierski:2005fc},
\begin{equation}
\label{gen_alg}
	[ \hat{x}_\mu \, \hat{x}_\nu ] ~~=~~ \frac i {\kappa^2} \theta_{\mu\nu} (\kappa \hat{x})~.
\end{equation}
Here $ \kappa $ is the scale parameter of the theory, while $ \theta_{\mu\nu} $ in general is a function
of coordinates:
\begin{equation}
\label{theta_series}
	\theta_{\mu\nu}(\hat{y}) ~~=~~ \theta_{\mu\nu}^{(0)} ~+~ \theta_{\mu\nu}^{(1)}{}^{\rho}\hat{y}_\rho ~+~
					\theta_{\mu\nu}^{(2)}{}^{\rho\sigma}\hat{y}_\rho\hat{y}_\sigma ~+~ \dots~.
\end{equation}
A typical feature of noncommutative field theories is the violation of ordinary Lorentz invariance.
Although the algebra \eqref{gen_alg} is invariant under quantum Poincar\'{e} transformations, it explicitly
breaks the usual Lorentz symmetry at the effective field theory level. 
At that point the noncommutativity comes in contact with phenomenology and observations. 

The most commonly encountered example of a noncommutative theory is the canonical noncommutativity \cite{Douglas:2001ba}, 
where only the constant part in the right hand side of \eqref{theta_series} is nonzero:
\begin{equation}
\label{theta_alg}
	[ \hat{x}_\mu \, \hat{x}_\nu ] ~~=~~ i\,\theta_{\mu\nu}~.
\end{equation}
A generic feature of a noncommutative field theory manifest here is 
{\it nonlocality}, which may be inherited  from
{\em e.g.} string theory.
Mild nonlocality at small length scales was thought to soften the problem of singular local field operators.
Indeed, the algebra \eqref{theta_alg} acts in a role of the ``Heisenberg relations'' for the coordinates,
preventing the constituent fields from being measured at the same position, 
and thus rendering all operators nonlocal. A commonly used tool in the analysis of 
noncommutative theories is the Seiberg-Witten map 
\cite{Seiberg:1999vs} which allows one to map a given theory to an 
effective field theory on a regular space-time 
by   means of the $*$-product. Moreover, it is likely that the energy scale associated 
with noncommutativity is much larger than the experimentally accessible 
scales, so that at tree level one can expand the $*$-extended theory 
in the series of local Lorentz-noninvariant
operators of increasing dimension. Symbolically, we can represent this procedure
by the following set of transformations:
\begin{equation}
\label{expansion-theta}
\int \hat{\mc{L}}_{\rm NC} d^4 \hat{x} \to \int \mc{L}(*) d^4 x = S_{\rm inv} ~+~ 
\int \theta^{\mu\nu}O_{\mu\nu} d^4x ~+~ \dots~,
\end{equation}
where $\int \hat{\mc{L}}_{\rm NC} d\hat{x} $ is the original action on the noncommutative coordinates,
$\mc{L}(*)$ is the $*$-product extended Lagrangian in ordinary space, and $O_{\mu\nu}$ are the 
effective Lorentz-breaking operators that involve three or more fields. In case of the canonical 
noncommutativity, $O_{\mu\nu}$ is the collection of dimension six operators. $S_{\rm inv}$ is the original 
Lorentz-invariant action of the commutative field theory, which for the phenomenology-oriented 
applications must contain the Standard Model (SM).

The major problem for $ \theta $-noncommutative theories start when one tries to 
make a connection with phenomenology. In an interacting theory loop effects
induce corrections to the two-point functions that are absent in the initial theory. 
These corrections are of course UV-sensitive \cite{Minwalla:1999px}.
But even at the tree level one can show that any 
compound system such as a {\em free} nucleon, will receive  $\theta$-dependent  corrections 
that couple the angular momentum to the external direction specified by $ \theta_{\mu\nu} $
\cite{Mocioiu:2000ip}.
The absence of such interactions in the effective Hamiltonian for a 
nucleon has been tested experimentally with exquisite accuracy, 
far better than GeV$^2M_{\rm Pl}^{-1}$ \cite{Bear:2000cd,Berglund}. 
This puts the limit on the noncommutativity scale, $\sim (|\theta_{\mu\nu}|)^{-1/2} $,
above $ 10^{14}~\GeV $ \cite{Mocioiu:2000ip,Anisimov:2001zc,Carlson:2001sw}.
It is also important to keep in mind that at any given order in $\theta$ there is 
an inherent ambiguity in defining the noncommutative Lagrangian, which is sensitive to 
{\em e.g.} the ordering 
of fields in $\mc{L}(*)$.  

The other frequently discussed example, which will be considered in 
this paper, is the $ \kappa $-Minkowski noncommutativity
\cite{Freidel:2006gc,Freidel:2007hk,Freidel:2007yu,Meljanac:2007xb,Rim:2008tm},
where only the linear term in the expansion \eqref{theta_series} is retained:
\begin{equation}
\label{space}
	[ \hat{x}_0\,, \, \hat{x}_i ] ~~=~~ \frac{i}{\kappa} \hat{x}_i~.
\end{equation}
In this case $ \theta_{\mu\nu}^{(1)}{}^\rho $ is of the special form
\begin{equation}
\label{def_C}
	\theta_{\mu\nu}^{(1)}{}^\rho ~~\equiv~~  C_{\mu\nu}^\rho ~~=~~ 
		\kappa^{-1}(a_\mu \delta_\nu^{\ \rho} - a_{\nu} \delta_\mu^{\ \rho})~,
		\qquad\qquad  a^{\mu} ~=~ ( 1, 0, 0, 0)~.
\end{equation}
Such a theory is often invoked as a possible candidate for the effective low energy 
description of quantum gravity \cite{Smolin:2004sx,Freidel:2006gc,AmelinoCamelia:2002wr}.
Leaving aside the issue of credibility of such claims, we would like to examine 
the phenomenology of this model, following the basic steps developed for
canonical noncommutativity. Should this model indeed be connected to quantum gravity, 
one would naturally expect $\kappa$ to be on the order of Planck mass.  

To define the effective theory, one has to explicitly specify the action of the $*$-product, {\it e.g.}
\cite{Dimitrijevic:2003wv}
\begin{eqnarray}
	\phi(x) * \psi(x) ~~=~~ \phi(x) \cdot \psi(x) ~+~ \frac{i}{2} C^{\mu\nu}_\lambda x^\lambda \p_\mu \phi(x) \p_\nu \psi(x)
			~+~ \dots ~.
\end{eqnarray}
It is possible that $ \kappa $-Minkowski theories introduce ultraviolet 
modifications to the dispersion relations for elementary particles. 
In the literature, there is hardly any unity on this issue: 
different formulations of the $ \kappa $-Minkowski theories 
have led to different formulations of effective Lagrangians, and in particular:
\begin{itemize}
\item Lukierski {\it et al.} \cite{Lukierski:1991pn}  \\[-5mm]
\begin{equation}
\label{luk}
	\mc{L} ~~=~~ \ha\, \phi \left\lgroup\,
			\Box ~+~ m^2 ~+~ \frac{\p_t^4}{\wt{\kappa}^2} \,\right\rgroup \phi~,
\end{equation}
\item Dimitrijevic {\it et al.} \cite{Dimitrijevic:2003wv} \\[-5mm]
\begin{equation}
\label{wess}
	\mc{L} ~~=~~ \ha\, \phi \left\lgroup\,  \Box ~+~ m^2 ~-~ \frac{\p_t^2}{\wt{\kappa}^2} \Box 
				\,\right\rgroup \phi~,
\end{equation}
\item Freidel {\it et al.} \cite{Freidel:2006gc} \\[-5mm]
\begin{equation}
\label{freidel}
	\mc{L} ~~=~~ (\p_\mu\phi)^\dag \sqrt{1 ~+~ \Box/\kappa^2}\, (\p_\mu\phi) ~~+~~
		     m\, \phi^\dag \sqrt{1 ~+~ \Box/\kappa^2}\; \phi~.
\end{equation}
\end{itemize} 
Here $ \wt{\kappa} $ is introduced to absorb numerical factors of order one.
In the effective Lagrangian one is allowed to use the equation of motion, so that in the 
second example the corresponding correction can be reduced to $\sim\kappa^{-2}m^2 E^2$, while
the last example turns out to be completely Lorentz-invariant  $\sim\kappa^{-2}m^4$.
These examples employ the scalar fields, which bears no relevance for phenomenology. 
However, should the approach of Ref. \cite{Lukierski:1991pn} be applicable 
to quarks and leptons, the existence of cosmic rays with highest energy would 
imply the sensitivity to $ \kappa$ possibly as high as the 
Planck scale \cite{Gagnon:2004xh}, but would not be able to probe $\kappa> M_{\rm Pl}$ .

It is clear, however, that in order to have a maximum  sensitivity to $\kappa$
one should exploit {\em  linear} order in $\kappa^{-1}$, which would 
correspond to operators of dimension five in the effective Lagrangian.
It is very well known that dimension five Lorentz-noninvariant 
operators are limited much better than the inverse Planck mass 
\cite{Gleiser:2001rm,Sudarsky:2002ue,Myers:2003fd,Gagnon:2004xh,Jacobson:2005bg,
GrootNibbelink:2004za,Bolokhov:2005cj,Bolokhov:2007yc,Bertolami:2004bf},
and the properties of all effective operators at this dimension are well understood.
For $\kappa$-Minkowski noncommutative gauge theories such dimension five operators have been derived in 
Refs. \cite{Dimitrijevic:2003pn,Dimitrijevic:2005xw}, and can be generalized to the full SM. 
Following \cite{Dimitrijevic:2003wv,Dimitrijevic:2003pn,Dimitrijevic:2005xw}, we take 
the effective Lagrangian
\begin{eqnarray}
\label{expansion-kappa}
	\mc{L}_{\rm eff}(*) ~~=~~ \mc{L}_{\rm inv} ~+~ \frac{a^\mu}{\kappa}\,O_\mu ~+~ \dots~,
\end{eqnarray}
to be the {\em only} source of $\kappa$-dependence in the theory
and explore the phenomenological consequence of such construction. 
Breaking of Lorentz invariance by $a_\mu$ means that 
specifying $ \kappa $ does not define the theory completely, 
as there remains a residual sensitivity to the orientation
of $a_\mu$. 

While there could be a symmetry reason prohibiting the emergence of the linear terms in $ a^\mu $, 
we argue that in the $\kappa$-SM such higher-dimensional terms are expected to arise.
As in the case of canonical noncommutativity, effective Lagrangian (\ref{expansion-kappa}) 
introduces the coupling of nucleon spin with the external direction defined by $ a^{\mu} $.
Strong existing bounds on such interactions push the scale of noncommutativity 4-5 orders of 
magnitude above the Planck scale, posing a serious difficulty for interpreting 
the $ \kappa $-Minkowski field theory as the low energy theory of quantum gravity.


\section{$\kappa$QED and $\kappa$SM}
We take as the starting point the gauge theory on $\kappa$-Minkowski spacetime introduced in
\cite{Dimitrijevic:2003wv,Dimitrijevic:2003pn,Dimitrijevic:2005xw}.
As customary for non-commutative theories, it possesses a degree of ambiguity related
to the fact that the Seiberg-Witten map is not unique \cite{Barnich:2002pb}.
In particular, a number of $x$-dependent terms are present in the Lagrangian.
It was shown by the authors of \cite{Dimitrijevic:2005xw} that a U(1) theory has
a set of free parameters by a suitable choice of which the ambiguous terms can be set to zero. 
We make this choice for the matter of convenience only, as this will allow us to
concentrate in more detail on the interactions that purpose the most immediate
phenomenological interest.
To the first order in the deformation parameter \cite{Dimitrijevic:2005xw}, 
the Lagrangian of U(1) gauge theory is given by
\begin{align}
\label{qed}
\mc{L} ~~=~~  \ov{\psi} ( i\gamma^\mu \md_\mu ~-~ m ) \psi ~-~ \frac{1}{4}F_{\mu\nu}F^{\mu\nu}
	~-~ \frac{1}{4} C_\lambda^{\rho\sigma} 
	( \ov{\psi} \gamma_\rho \md_\sigma \md^\lambda \psi ~+~
		\ov{\md_\sigma \md^\lambda \psi} \gamma_\rho \psi )~.
\end{align}
Here $ C^{\mu\nu}_\rho $ and $ a^\mu $ are as in \eqref{def_C}.
The latter two terms in the brackets are dimension five interactions which describe the deviation of the theory 
from regular U(1) QED. 
Since all dimension five Lorentz-violating operators in QED were classified in 
\cite{Bolokhov:2007yc}, one should expect the operators in \eqref{qed} to be a specific 
realization of such operators.
To show that explicitly, we transform the Lorentz-noninvariant piece in \eqref{qed} 
using the equations of motion in the zeroth order in $\kappa$, which is allowed as long as 
we are satisfied with $O(\kappa^{-1})$ accuracy. Thus the equivalent form of (\ref{qed}) is given by
\begin{equation}
\label{red_qed}
\kappa~\mc{L}_{\rm QED} ~~=~~  e\, a^\mu \ov{\psi} \wt{F}_{\mu\nu} \gamma^\nu \gamma^5 \psi~, 
\end{equation}
where we have also substituted $C^{\rho\sigma}_\lambda $ in terms of $a^\mu$, and $\wt{F}_{\mu\nu}
=\frac12\epsilon_{\mu\nu\alpha\beta}F^{\alpha\beta}$. 
As one can readily see, the Lorentz violation disappears for a free fermion ($F_{\mu\nu}=0$).
Operator (\ref{red_qed}) was indeed encountered in the general analysis of Ref. \cite{Bolokhov:2007yc}.

This operator is $CPT$ and $ C $-odd, and its zeroth component is even under $ T $-parity. 
These discrete symmetries therefore allow \eqref{red_qed} to transmute at 
the loop level to the electromagnetic current operator
$O(\Lambda_{\rm UV}^2)\times \ov{\psi} \gamma^\mu \psi $.
However, this is inconsequential, since an additive contribution to the electromagnetic
current can be absorbed into the electromagnetic potential.

A much reacher structure emerges from the Standard Model, where parity is violated and the operators with 
the properties of the axial current are unavoidable. 
The interaction \eqref{qed} is uniquely generalized to include the $ U(1) $, $ SU(2) $ and the $ SU(3) $ gauge field
strengths via the covariant derivative
\[
	\md_\mu ~~=~~ \p_\mu ~+~ i g' B_\mu ~+~ i g W_\mu^a \frac{\tau^2}{2} ~+~ i g_3 G_\mu^a t^a~.
\]
Applying this explicitly for quarks and using the equations of motion we obtain the following combination:
\begin{align}
\notag
\kappa~\mc{L}_{\rm SM} ~~=~~ & ~-\, Y_Q\, g'\, a^\mu \ov{Q} \wt{B}_{\mu\nu} \gamma^\nu Q 
		     ~-~ g\, a^\mu \ov{Q} \wt{W}_{\mu\nu}^a \frac{\tau^a}{2} \gamma^\nu Q 
		     ~-~ g_3\, a^\mu \ov{Q} \wt{G}_{\mu\nu}^a t^a \gamma^\nu Q \\
\label{sm}
	     & ~+~ Y_U\, g'\, a^\mu \ov{U} \wt{B}_{\mu\nu} \gamma^\nu U ~+~ g_3\, a^\mu \ov{U} \wt{G}_{\mu\nu}^a t^a \gamma^\nu U 
	      ~+~ Y_D\, g'\, a^\mu \ov{D} \wt{B}_{\mu\nu} \gamma^\nu D ~+~ g_3\, a^\mu \ov{D} \wt{G}_{\mu\nu}^a t^a \gamma^\nu D
			~,
\end{align}
where $ Y_Q $, $ Y_U $, $ Y_D $ denote the corresponding hypercharges. 
These Lorentz-violating operators are "soft" \cite{Bolokhov:2007yc}, that is their size does not grow with the 
energy of the fermion. 
At low energies we can restrict our analysis only to the color field strength operators and the operators involving $ B_{\mu\nu} $
and $ W_{\mu\nu}^3 $.
The former ones collect into
\[
	g_3\, \ov{q}\, \wt{G}_{\mu\nu}^a t^a \gamma^\nu \gamma^5 q~.
\]
The combination of \eqref{red_qed} for $d$ and $s$ quark could potentially induce the $CPT$-odd mass shift in the 
sector of neutral $K$-mesons.
The other operators combine into 
\begin{align}
\notag
	\kappa \mc{L}_{\rm 1~GeV} ~~=~~
	\ha( Y_U - Y_Q - 1 )\, e\, a^\mu \wt{F}_{\mu\nu} \ov{u} \gamma^\nu u ~~+~~
	\ha( Y_D - Y_Q + 1 )\, e\, a^\mu \wt{F}_{\mu\nu} \ov{d} \gamma^\nu d ~~+~~ \\
\label{sm_red}
	\ha( Y_U + Y_Q + 1 )\, e\, a^\mu \wt{F}_{\mu\nu} \ov{u} \gamma^\nu \gamma^5 u ~~+~~
	\ha( Y_D + Y_Q - 1 )\, e\, a^\mu \wt{F}_{\mu\nu} \ov{d} \gamma^\nu \gamma^5 d~,
\end{align}
where we only listed the terms which include the electromagnetic field strength.
The second line in this expression is the complete analogue of \eqref{red_qed}.
The first line in \eqref{sm_red}, 
\begin{equation}
\label{T_odd}
	\kappa \mc{L}_{\rm T-odd} ~~=~~ C_u\, a^\mu \wt{F}_{\mu\nu} \ov{u} \gamma^\nu u ~~+~~
				 C_d\, a^\mu \wt{F}_{\mu\nu} \ov{d} \gamma^\nu d,
\end{equation}
where $C_{u(d)}$ are introduced for concision, contains the coupling of $a^0$ to operators that are $ T $-odd, $P,C$-even,  
and thus have the same properties as the axial vector
current \cite{Bolokhov:2006yx}:
%
\begin{equation}
\label{effectiveK}
	\mc{L}_{\rm axial} ~~=~~ b^\mu\, \ov{\psi} \gamma_\mu \gamma^5 \psi~,
\end{equation}
which is part of the Colladay-Kostelecky effective Lagrangian 
\cite{Colladay:1996iz}.
At low energies, we expect it to induce an interaction of the nuclear spin to the external 
direction specified by $ \vec{b} $.
This type of interactions is strongly constrained by experiment \cite{Bear:2000cd}, and it is therefore
instructive to use the operators \eqref{T_odd}
to transfer these experimental limits on the scale of noncommutativity $ \kappa $.

Of course, our analysis is prone to the same conceptual difficulty as the canonical noncommutativity 
due to the UV-divergent mixing arising  at the loop level \cite{Bolokhov:2007yc}. The diagrams that include these interactions 
produce quadratic divergencies, and the effective theory is then highly sensitive to the cut-off scale $ \LUV $
\begin{eqnarray}
	a^\mu \ov{\psi} \wt{F}_{\mu\nu}\gamma^\nu \psi  \qquad \Rightarrow \qquad \LUV^2\, a^\mu \ov{\psi} \gamma_\mu \gamma^5 \psi~.
\end{eqnarray}
In principle, the presence of the ultraviolet scale in the effective Lagrangian, {\em e.g.} supersymmetry breaking scale,
would imply more stringent constraints on the parameter of noncommutativity. 
We choose 
to leave the issue of the radiative corrections aside and instead 
estimate the effective coupling of nucleon axial vector current with $a^\mu$ \eqref{effectiveK} that results from
nucleon compositeness. 
To do so, we need to find the order of magnitude of the coupling strength $ b^\mu $,
which is related to the parameter of noncommutativity via a QCD matrix element:
\begin{equation}
	b^\mu ~~\sim~~ a^\mu \frac{\alpha_{\rm QED}}{4\pi\kappa} \Lambda_{\rm hadr}^2~,
\end{equation}
where the fine structure constant originates from the internal photon exchange between quarks inside the nucleon, and 
$\Lambda_{\rm hadr} \sim m_n$ is the characteristic hadronic energy scale required by dimension. 

The easiest way to obtain an estimate for this coefficient is to use the Vector Meson Dominance Model.
We assume that the nucleon electromagnetic interactions  are mediated by $ \rho,\omega $-mesons.
 LV coupling \eqref{T_odd} leads to the additional photon-vector meson mixing of the 
form $a^\mu \wt{F}{}_{\mu\nu} V^\nu$.  
Calculating a simple loop with the insertion of $CPT$-odd $V-\gamma$ interaction, 
and specializing it to the case of the neutron, for which the experimental constraints are the 
most stringent, we obtain the neutron axial form factor as
\begin{eqnarray}
	\langle n|\,  
	C_u a^\mu \wt{F}_{\mu\nu} \,
		\ov{u} \gamma^\nu u
	~+~
	C_d a^\mu \wt{F}_{\mu\nu} \,
		\ov{d} \gamma^\nu d
	\, |n \rangle ~~=~~
	-\,\frac{3}{\sqrt{2}\,32\pi^2}\log \frac{m_N}{m_\rho}\, e \mu_n ( \mu_p - \mu_n ) m_\rho^4\,
			a^\mu \cdot \ov{n} \gamma^\mu \gamma^5 n~ \nonumber\\
			= 3 \times 10^{-5} {\rm GeV}^2 \times a^\mu \cdot \ov{n} \gamma^\mu \gamma^5 n~,
\end{eqnarray}
where $\mu_p, \mu_n$ are the anomalous magnetic moment of nucleons, and only the log-divergent term is retained. 

Armed with these estimates, we are ready to translate the experimental limits on $b^i$(neutron),
$|b_i|< 10^{-31}$GeV into the sensitivity to $\kappa$.
Given the strength of the constraints on the spatial components of $ b^i < 10^{-31}~\GeV $, we
obtain a limit on $ \kappa^{-1}|\vec{a}| $ better than $ 10^{-27}~\GeV^{-1} $. If one takes the point of view
that the spatial components of $ a^\mu $ are not part of the original $ \kappa $-Minkowski theory, then the 
former will still be induced by the motion of the Earth in the background frame in which the theory is defined
({\it e.g.} the CMB frame). 
The constraint then weakens by some 3 orders of magnitude, and a rather conservative bound of 
\begin{equation}
\label{bound}
	\kappa ~~>~~ 10^{23-24}~\GeV
\end{equation}
is obtained. This is the main result of our analysis. 
Bound \eqref{bound} is on par with all other bounds for the so-called "soft" dimension five
Lorentz breaking operators \cite{Bolokhov:2007yc}.  


\section{Discussion}

In this paper, we have considered the extension of gauge theories consistent with $\kappa$-Minkowski
symmetry in the first order in $\kappa^{-1}$, developed in \cite{Dimitrijevic:2003pn,Dimitrijevic:2005xw}, 
and applied it to the Standard Model. The outcome, Eq. \eqref{bound}, may look devastating 
for the attempts to promote $\kappa$-Minkowski field theory into the low-energy theory of 
quantum gravity, as the constraint on $\kappa$ is several orders of magnitude stronger than the Planck mass scale.
We would like to point out that the 
strength of the constraint is the direct consequence of linear $a^\mu$ dependence in 
\eqref{expansion-kappa}, which is the {\em only} source
of $1/\kappa$ terms, and is at the core of the whole approach in Refs. \cite{Dimitrijevic:2003pn,Dimitrijevic:2005xw}.
We believe that this is the correct approach that has a well-defined operational meaning. Effective action 
\eqref{expansion-kappa} can be applied to study {\em any} observable at any given order in inverse $\kappa$. 
Moreover, we think that other approaches where some extra sources of $\kappa$ are invoked 
{\em e.g.} inside the wave functions \cite{Arzano:2007gr,Arzano:2007ef,Arzano:2007qp} 
lack clear operational meaning to the same extent that 
Eq. (\ref{expansion-kappa}) does. In  any event, the expansion over
$\kappa$ is always possible even for the wave function, and we expect that the 
resulting (possibly $x^\mu$-dependent) $\kappa^{-1}$-terms can be again cast as terms in the 
effective action. Whether or not this would eliminate explicit $a^\mu$-dependence in 
\eqref{expansion-kappa} is another issue, and it falls outside the scope of the present paper.  

Is it possible to remove the linear dependence on $\kappa^{-1}$ in QED and the SM
by exploiting the ambiguities inherent in the noncommutative field theories? 
For QED, the recipe is very simple.
Viewed from the point of undeformed Lorentz symmetry, parameter of deformation $\kappa^{-1} a^\mu$ is a 
Lorentz vector with the properties of the vector current under the discrete transformations. 
One can exploit the $C$-odd properties of \eqref{red_qed}, and have $\mc{L}_{\rm QED}(*) \to \frac12 \mc{L}_{\rm QED}(*)
+ \frac12 C(\mc{L}_{\rm QED}(*))$, which would leave main QED physics unchanged but eliminate 
linear in $1/\kappa$ correction. For the SM this would not work: the unperturbed Lagrangian of the SM 
contains $C$-odd terms, which this recipe will remove. In canonical noncommutativity  there is, however, a very efficient way of 
removing the linear dependence in $\kappa^{-1}$ by combining terms with different field ordering in the
$*$-modified Lagrangian. This method may work for $\kappa$-Minkowski field theories as well, 
although in our opinion such engineering of $\mc{L}(*)$ would also not look particularly natural.

With our approach we observe the same phenomenon in $\kappa$-Minkowski effective field theory as in the 
case of the canonical noncommutativity. If at the level of constituents there is a symmetry that removes 
$\kappa$-dependence in the two-point functions and leaves it only in the interaction terms, 
{\em e.g.} Eqs. (\ref{qed}) and (\ref{red_qed}), at the level of composite systems (nucleons, nuclei, atoms etc)
this is no longer true and the modification of the Hamiltonian for free composite particles emerges (\ref{effectiveK}).
This is reminiscent of the "composition problem" that exists for all 
exotic theories with nonlinear dependence of dispersion relation on energy and momenta, for example, one does not expect that
 $E^3/M_{\rm Pl}$ corrections postulated for constituents would hold in exactly the same form for a composite system.




\section{Acknowledgments}
M.P. would like to acknowledge useful conversations with L. Freidel. 
The work of PAB was supported in part by the NSF Grant No. PHY-0554660. 
This work was supported in part by NSERC, Canada, and research at the Perimeter Institute
is supported in part by the Government of Canada 
through NSERC and by the Province of Ontario through MEDT.

\end{document}